\documentclass[prb,reprint,aps,superscriptaddress,longbibliography]{revtex4-2}

\usepackage[pdftex]{graphicx}

\usepackage{dcolumn}
\usepackage{bm}
\usepackage{amsmath}
\usepackage{amssymb}
\usepackage{ulem}
\usepackage{float}

\usepackage[caption=false]{subfig}
\graphicspath{ {img/} }

\usepackage{hyperref}
\usepackage[usenames,dvipsnames]{xcolor}
\hypersetup{colorlinks=true, 
linkcolor=Violet, 
citecolor=ForestGreen, 
filecolor=BrickRed, 
urlcolor=RoyalBlue, 
filebordercolor={.8 .8 1}, 
urlbordercolor={.8 .8 0}
}%

\usepackage{tikz}
\usetikzlibrary{positioning}

\newcommand{\<}{\langle}

\renewcommand{\>}{\rangle}
\renewcommand{\(}{\left(}
\renewcommand{\)}{\right)}
\renewcommand{\[}{\left[}
\renewcommand{\]}{\right]}

\renewcommand{\d}{\partial}

\newcommand{\mat}[1]{\ensuremath{\underline{#1}}}

\begin{document}

\title{Valence-bonds, spin liquids and unconventional criticality in a 1D Kondo insulator}

\author{Nai Chao Hu}
\email{naichao.hu@ugent.be}
\affiliation{Department of Physics and Astronomy, Ghent University, Krijgslaan 281, S9, 9000 Gent, Belgium}

\newcommand{\gscaep}[0]{Graduate School of China Academy of Engineering Physics, Beijing 100193, China}
\author{Rui-Zhen Huang}
\affiliation{\gscaep}

\author{Nick Bultinck}
\email{nick.bultinck@ugent.be}
\affiliation{Department of Physics and Astronomy, Ghent University, Krijgslaan 281, S9, 9000 Gent, Belgium}

\date{\today}

\begin{abstract}
We consider a one-dimensional multi-orbital Kondo lattice model and show that by tuning the kinetic energy of the itinerant electrons it is possible to stabilize Kondo insulators with non-trivial spin physics. In particular, depending on the size of the exchange coupling between the local moments, we find kinetic-energy-driven transitions between a featureless Kondo insulator and a valence-bond solid or a gapless spin liquid. We also provide evidence for an unconventional continuous phase transition between two featureless Kondo insulators distinguished by their quantum numbers under reflection symmetry.
\end{abstract}
\maketitle

\section{Introduction}
Kondo insulators acquire their incompressible character via a spin exchange coupling between itinerant charge carriers and local moments. Doping a Kondo insulator gives rise to a heavy Fermi liquid with exceptionally large effective masses for the low-energy quasi-particles~\cite{stewart1984heavy,gegenwart2008quantum,paschen2020quantum}. One of the defining features of both Kondo insulators and heavy Fermi liquids is that they violate the Luttinger sum rule which relates the local charge density to the Fermi surface volume, unless the charge density takes both the itinerant electrons and the local moments into account. More broadly, heavy fermion materials present a paradigmatic class of strongly-correlated electron systems, exhibiting rich phase diagrams including e.g. magnetic states, unconventional superconductivity, and non-Fermi liquid physics~\cite{stewart1984heavy,gegenwart2008quantum,paschen2020quantum,coleman2001fermi,senthil2004weak}.

The lattice models used to understand Kondo insulators and heavy Fermi liquids are typically studied when the itinerant electron bandwidth is the largest energy scale in the problem. The strength of Kondo screening of the local moments is then determined by the dimensionless number $J_K D(E_F)$, where $J_K$ is the exchange coupling between the local moments and the spin of the charge carriers, and $D(E_F)$ is the density of states at the Fermi energy. In this parameter regime, only the Bloch states near the Fermi energy participate in the Kondo screening, and hence global properties of the Bloch band (i.e. the topological properties) should not matter too much. On the other hand, for flat-band systems where the bandwidth becomes comparable to $J_K$, we can expect the global structure of the band to become important.

In this work we consider a one-dimensional model where the non-trivial global property of the itinerant electron band is its polarization, quantized at $1/2$ due to spatial reflection symmetry. In real space, this means that the Wannier centers of the itinerant electrons are located in between the unit cells. Upon decreasing the bandwidth we find a transition from the conventional featureless Kondo insulator to either a valence-bond solid or one of two types of gapless spin liquids. Exactly which state is obtained at small bandwidth depends on the value of the spin exchange coupling between the local moments. Given that for a completely trivial band the weak-coupling Kondo insulator (large bandwidth) is adiabatically connected to the strong-coupling Kondo insulator (small bandwidth), we thus find that the most trivial form of non-trivial band topology already drastically changes the Kondo insulator phase diagram.

When the exchange coupling between the local moments is much larger than the Kondo coupling between local moments and charge carriers we provide evidence that the featureless Kondo insulator survives for any non-zero bandwidth, but becomes critical when the band is flat. This critical point separates two types of featureless Kondo insulators with different quantum numbers under the spatial reflection symmetry. As explained in more detail below, we argue that this realizes a critical point between a trivial and a non-trivial fragile Kondo insulator.

Although the model studied in this work should be regarded for what it is, namely a theoretical toy model, many of the key ingredients such as flat bands, strong interactions and non-trivial topology, are present in moir\'e materials. In fact, Kondo lattice systems have already been experimentally realized in moir\'e systems~\cite{Zhao2023,Zhao2024}, and the interplay between topology and Kondo physics has been explored in detail for these materials~\cite{Dalal2021,Kumar2022,Song2022,Calugaru2023,Guerci2023,Guerci2024,Xie2024,Herzog2024,Singh2024}. The results obtained in this work could therefore serve as a further motivation for exploring the different types of Kondo lattice systems that can be engineered in moir\'e lattices.

\begin{figure}[th!]
    \centering
    \includegraphics[width=0.49\textwidth]{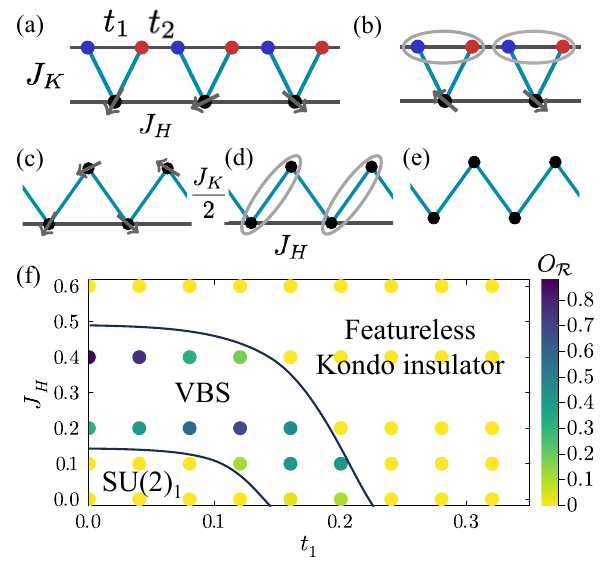}
    \caption{(a) Lattice structures, where black circles represent spin sites and colored circles represent electron sites. (b) Product state structure of special case 1 (see main text), where $c$ electrons are in bonding or anti-bonding state. (c) Effective spin model of special case 2 (again see main text).  (d) Schematic of the VBS state near $J_H = J_K/2$. (e) Effective Heisenberg chain when $J_H = 0$. (f) Reflection symmetry-breaking order parameter $O_{\mathcal{R}}$ defined in Eq. \eqref{defOR} (obtained without explicit SU(2) symmetry in the MPS) in the $J_H-t_1$ plane obtained with $t_2 = 1,\ J_K=0.6$.} 
    \label{fig:phases}
\end{figure}

\section{Multi-orbital Kondo model} 
We consider a one-dimensional lattice model with itinerant $c$-electrons and Mott-localized $f$-electrons giving rise to local moments. The unit cell contains one local moment, and two ($A$ and $B$ sublattice) sites for the $c$-electrons (see Fig.~\ref{fig:phases}(a)). The Kondo lattice Hamiltonian is given by
\begin{align}
   H =  & -t_1 \sum_{j,\sigma} c^\dagger_{j,A,\sigma}c_{j,B,\sigma} - t_2 \sum_{j,\sigma} c^\dagger_{j,B,\sigma}c_{j+1,A,\sigma} + h.c. \nonumber \\
   & +\frac{J_K}{2} \sum_{j} \mathbf{S}_j \cdot \left(c^\dagger_{j,A}\boldsymbol{\sigma}c_{j,A} + c^\dagger_{j,B}\boldsymbol{\sigma}c_{j,B,}  \right) \nonumber\\
   & + J_H \sum_j \mathbf{S}_j\cdot\mathbf{S}_{j+1}\,, \label{defH}
\end{align}
where $j$ labels the unit cells, $\mathbf{S}_j$
are the spin operators of the local $f$ moments, and the Kondo and Heisenberg couplings $J_K$ and $J_H$ are both positive. In the following, we focus on the specific case of quarter-filling for the $c$-electrons (i.e. one $c$-electron per unit cell), such that Kondo hybridization leads to incompressible states.

The model in Eq. \eqref{defH} contains four free parameters, leading to a three-dimensional phase space that can be parameterized by the ratios $J_H/J_K$, $t_1/t_2$, and $J_K/t_2$. The ratio $J_H/J_K$ influences the magnetic correlations, while $t_1/t_2$ governs the dispersion and structure of the Bloch wavefunctions in the partially filled $c$-electron band. The ratio $J_K/t_2$ interpolates between weak- and strong-coupling Kondo insulators. We will primarily be interested in how the ground state changes upon varying $J_H/J_K$ and $t_1/t_2$. In particular, for most of our analysis we will fix $J_K = 0.6$ and $t_2=1$, and vary both $J_H$ and $t_1$.

\subsection{Special case 1: $t_2=0$, $t_1\neq 0$}
To gain some first understanding of the physics realized by the Hamiltonian in Eq. \eqref{defH} we consider two special cases in parameter space. First, consider the case where $t_2=0$ and $t_1 \neq 0$. The $c$-electrons are now localized in the unit cell (see Fig.~\ref{fig:phases}(b)), and depending on the sign of $t_1$ occupy either bonding ($t_1>0$) or anti-bonding $(t_1<0)$ states between the $A$ and $B$ sublattice sites. Since fluctuations of the total charge in every unit cell are strictly zero, the $c$-electrons now also behave as local moments, which are anti-ferromagnetically coupled to the local $f$-moments. The latter form a spin-$1/2$ Heisenberg anti-ferromagnet, which is known to be unstable to the formation of a spin gap upon exchange-coupling an additional spin-$1/2$ in every unit cell~\cite{Haldane1983,BozSpin}. Because $J_K>0$, the spin-gapped phase is trivial (with $J_K<0$ we would obtain the symmetry-protected topological Haldane phase). This is especially clear in the limit $J_H \rightarrow 0$, in which case the ground state becomes a simple product state:
\begin{eqnarray}
|\psi_+\rangle & = & \bigotimes_j \frac{1}{\sqrt{2}} \left(c_{j,+,\uparrow}^\dagger c^\dagger_{j,f,\downarrow} - c^\dagger_{j,+,\downarrow} c^\dagger_{j,f,\uparrow} \right) |0\rangle \,,\;\;(t_1>0) \nonumber\\
|\psi_-\rangle & = & \bigotimes_j \frac{1}{\sqrt{2}} \left(c_{j,-,\uparrow}^\dagger c^\dagger_{j,f,\downarrow} - c^\dagger_{j,-,\downarrow} c^\dagger_{j,f,\uparrow} \right) |0\rangle\,, \;\;(t_1<0) \nonumber
\end{eqnarray}
where $c^\dagger_{j,\pm,\sigma} = \frac{1}{\sqrt{2}}(c^\dagger_{j,A,\sigma} \pm c^\dagger_{j,B,\sigma})$ creates a $c$-electron in a local bonding or anti-bonding state in unit cell $j$, and $c^\dagger_{j,f,\sigma}$ creates an $f$-electron in unit cell $j$. The above product states are the standard strong coupling Kondo-insulator ground states of the conventional Kondo lattice model. The two states $|\psi_+\rangle$ are $|\psi_-\rangle$ are both trivial, but are nevertheless sharply distinct due to the spatial reflection symmetry of the model. In particular, under a spatial reflection centered on an $f$-moment in a system with $L$ unit cells we get
\begin{eqnarray}
\mathcal{R}|\psi_+\rangle & = & |\psi_+\rangle \\
\mathcal{R}|\psi_-\rangle & = & (-1)^L|\psi_-\rangle\,, \label{Rmin}
\end{eqnarray}
which shows that the two product states have a different reflection quantum number (per unit cell), and hence cannot be connected without crossing a phase transition if the reflection symmetry of the Hamiltonian is maintained. In fact, from Eq. \eqref{Rmin} we know that $|\psi_-\rangle$ cannot be adiabatically connected to any band insulator if reflection symmetry is preserved, as it was shown in Ref.~\cite{fragileMott} that band insulators necessarily have trivial quantum numbers under point group symmetries. For this reason, we could call $|\psi_-\rangle$ a non-trivial {\it fragile Kondo insulator}.

\subsection{Special case 2: $t_1=0$, $t_2>0$}
Next, consider the case with $t_1 = 0,\ t_2 > 0$. Now the $c$-electrons are localized in bonding states on the $t_2$ bonds, i.e. the pairs of sites $\{(j,B),(j+1,A)\}$, and again effectively behave as spins. However, now the geometry of the resulting spin model is crucially different from the previous case with $t_2=0$, because the $c$-electrons form local moments which are located exactly in between the $f$-moments (see Fig.~\ref{fig:phases}(c)). The resulting spin model has been studied in detail before, and goes under the name of the {\it sawtooth spin chain} \cite{Nakamara1996,Sen1996,Derzhko2020,Schulenburg2002,Blundell2003,Richter2004,Richter2005,sawtooth,Jiang2015,Hutak2023}. Crucially, even though the sawtooth spin chain has integer spin per unit cell, there is a Lieb-Schultz-Mattis (LSM) obstruction to a gapped ground state which respects both the spin SU(2) and the spatial reflection symmetry. In our case, the $f$-moments couple to the $c$-moments with an exchange coupling of $J_K/2$, where the factor of $1/2$ comes from the fact that the $f$-moment at site $j$ couples to the $c$-electron on bond $\{(j,B),(j+1,A)\}$ only when it sits on the $(j,B)$ site, which it does with probability $1/2$. This results in a sawtooth-spin chain where the basal $f$-moments interact with exchange coupling $J_H$, and couple to the apex $c$-moments with strength $J_K/2$. 

When $J_H=0$, the chain forms a spin-$1/2$ anti-ferromagnetic Heisenberg chain of alternating $f$- and $c$-moments, with exchange coupling $J_K/2$ (see Fig.~\ref{fig:phases}(e)). This model is well-known to be gapless, and is described by the SU(2)$_1$ Conformal Field Theory (CFT). 

Next, $J_H=J_K/2$ corresponds to the solvable Majumdar-Ghosh point \cite{majumdar-ghosh} of the model, with two degenerate Valence-Bond Solid (VBS) ground states with zero correlation length. The two VBS states correspond to putting the $f$-moments in unit cell $j$ in a singlet state with the $c$-moments in either bond $\{(j,B),(j+1,A)\}$ or $\{(j-1,B),(j,A)\}$ (see Fig.~\ref{fig:phases}(d)). The two VBS states are related by a spatial reflection, meaning that this symmetry is spontaneously broken. 

Lastly, the limit $J_H \gg J_K$ is highly non-trivial. In this case the apical spins act as perturbations to the basal Heisenberg chain.  
In Ref. \cite{sawtooth}, the $J_H \gg J_K$ regime was studied numerically with DMRG, where it was found that the apical $c$ chain exhibits algebraic magnetic spiral correlations with incommensurate period, while the basal $f$-chain retains its algebraic anti-ferromagnetic correlations. In contrast, solutions derived from Luttinger-Tisza-like methods impose identical periodicity across sublattices.

The authors of Refs. \cite{sawtooth,Jiang2015,Hutak2023} have mapped out the complete phase diagram of the sawtooth spin chain. They have found that (1) the transition from the gapless SU(1)$_1$ spin liquid to the VBS phase occurs at finite $J_H/J_K \sim 0.24$, and (2) the VBS phase transitions into the incommensurate spiral phase at $J_H/J_K \sim 0.75$. For large $J_H/J_K$ the magnetic correlations in the apex $c$-moments become commensurate and correspond to $90^\circ$ spiral correlations \cite{sawtooth}. As expected from the LSM obstruction, all phases either break the reflection symmetry (VBS), or are gapless (SU(2)$_1$ and spiral spin liquid).

\begin{figure*}[t]
    \centering
    \includegraphics[width=\linewidth]{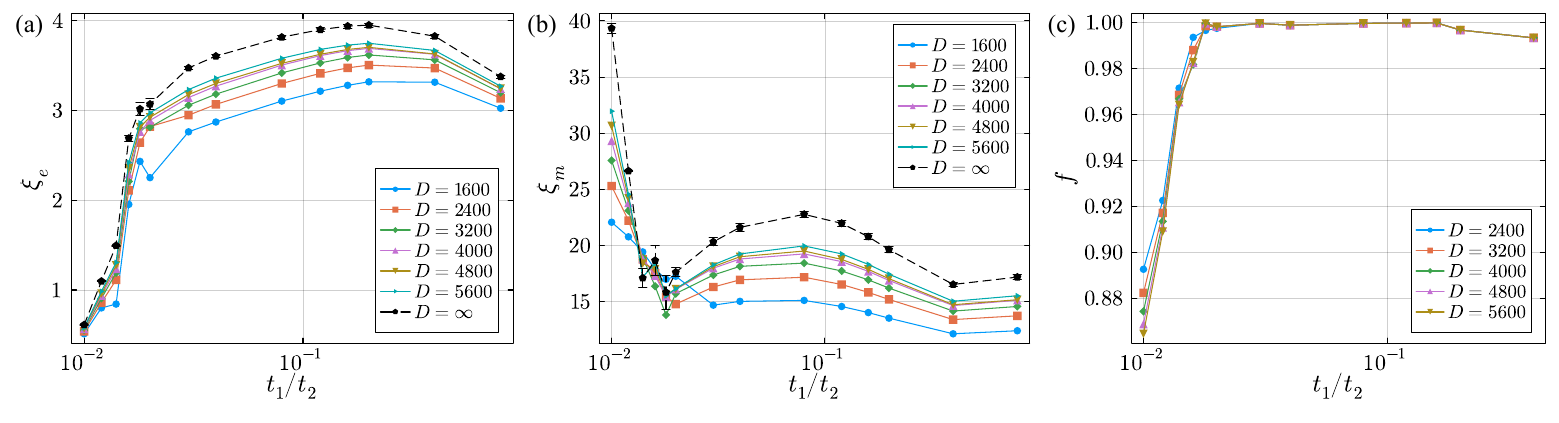}
    \caption{Numerical evidence of a gapped phase in both the charge and spin sectors at large $J_H$ ($J_H/J_K=2$). (a)-(b) Electron and spin correlation lengths $\xi_{e/m}$ as a function of $t_1/t_2$ for different bond dimensions $D$. The black points are obtained from a linear extrapolation of $1/\xi = m/D + 1/\xi_{\infty}$, where $m$ and $\xi_{\infty}$ are fitting parameters. (c) Fidelity density $f$ defined in Eq.~\eqref{deff}, obtained from pairs of MPS with neighboring $t_1$ values. }
    \label{fig:gap}
\end{figure*}

We note that in the Kondo lattice model we expect the SU(2)$_1$ and VBS phases to be realized as incompressible states, because adding or removing a $c$-electron corresponds to removing an apex spin in the sawtooth spin chain. In the SU(2)$_1$ and VBS states this clearly leads to a loss of negative exchange energy (e.g. at the Majumdar-Ghosh point it corresponds to breaking a singlet). The spiral spin liquid, on the other hand, we will show below to be realized as a compressible phase transition line in the Kondo lattice model. To understand the physics of this phase transition, let us first point out the important role played by the reflection symmetry $\mathcal{R}$. It is this symmetry which distinguishes the two classes of trivial Kondo insulators adiabatically connected to either $|\psi_+\rangle$ or $|\psi_-\rangle$, and for the sawtooth spin chain obtained at $t_1=0$ it gives rise to the LSM obstruction for featureless gapped ground states. We will show below that both these aspects of the reflection symmetry are intimately connected. In particular, our numerical simulations reveal that the gapless spiral spin liquid of the sawtooth spin chain can be interpreted as a critical point between the two different classes of Kondo insulators. This critical point is special for two reasons. First, the gapped phases on either side of this critical line are symmetric and topologically trivial -- they are two distinct fragile Kondo insulators. This places the transition outside the usual Landau paradigm for second-order phase transitions. Second, despite its seemingly benign appearance, the critical theory is not described by a CFT, as we will argue with several pieces of numerical evidence.

\section{Numerical results} 
We numerically compute the phase diagram of the Hamiltonian Eq.~\eqref{defH} with the variational uniform matrix product state (VUMPS) algorithm \cite{vumps, van2024mpskit, haegeman2024tensorkit} and conjugate gradient when VUMPS convergence is slow \cite{conjugateGradient}. Throughout this work we denote the bond dimension of the MPS used in our simulations as $D$. 

\subsection{Valence-bond solid and SU(2)$_1$ spin liquid}
In Fig.~\ref{fig:phases}(f) we show the value of the reflection symmetry-breaking order parameter
\begin{equation}
O_{\mathcal{R}} = \Big\langle \mathbf{S}_{j}\cdot\left(c^\dagger_{j,B}\frac{\boldsymbol{\sigma}}{2}c_{j,B} -c^\dagger_{j,A}\frac{\boldsymbol{\sigma}}{2}c_{j,A} \right)\Big\rangle \label{defOR}
\end{equation}
for different values of $t_1$ and $J_H$ (with fixed $J_K=0.6$ and $t_2=1$). To obtain this phase diagram we have not imposed SU(2) symmetry on the MPS, because the LSM obstruction implies that an SU(2)-symmetric matrix product state (MPS) with finite bond dimension must (weakly) break the reflection symmetry. So by not imposing SU(2) we expect to reduce the finite bond dimension-induced reflection symmetry breaking, and as result, obtain a more accurate order parameter value. Also note that we only compute the $t_1 > 0$ region of the phase diagram. This is because for an even number of unit cells (or directly in the thermodynamic limit as in our simulations) there exists a unitary transformation 
\begin{equation}
V = \prod_j (-1)^{\sum_{\sigma}c^\dagger_{2j,A,\sigma}c_{2j,A,\sigma} + c^\dagger_{2j+1,B,\sigma}c_{2j+1,B,\sigma}} \label{defU}
\end{equation}
which changes the sign of $t_1$: $V^\dagger H(t_1) V = H(-t_1)$. Consequently, the phase diagram is symmetric about $t_1 = 0$. The region with reflection symmetry breaking obtained at finite $t_1$ matches the VBS phase in the sawtooth spin chain on the $t_1=0$ axis. We also see that the SU(2)$_1$ spin liquid of the sawtooth spin chain at small $J_H$ extends into a finite region in the $(J_H,t_1)$ plane where the reflection symmetry-breaking order parameter is zero. To further characterize this region, we performed calculations enforcing full spin SU(2) symmetry on the MPS and extract the spin correlation length $\xi_m$ of the ground state MPS from the largest transfer matrix eigenvalue in the charge-$0$ and spin-$1$ symmetry sector. We also calculated the entanglement entropy $S$, which is found to follow the universal CFT relation $S = \frac{c}{6}\log\xi_m + \text{const}$, with a central charge $c = 1$. This confirms that the SU(2)$_1$ spin liquid is realized in an extended parameter region at small $J_H $ and $t_1$. 

Starting inside the SU(2)$_1$ spin liquid region, as either $J_H$ or $t_1$ is sufficiently increased reflection symmetry breaking sets in. In this symmetry-broken phase, we find that both the spin correlation length $\xi_m$ and the electron correlation length $\xi_e$ (charge-$1$, spin-$1/2$ sector) quickly saturate as a function of bond dimension $D$, consistent with an incompressible and spin-gapped VBS state. In the range $0.15 \lesssim t_1 \lesssim 0.2$ the VBS phase extends all the way down to the $J_H=0$ axis. This shows that the SU(2)$_1$ spin liquid is shielded from the featureless Kondo insulator at large $t_1$ by an intermediate VBS region. The transition from the VBS state to the featureless Kondo insulator appears rather abrupt in our VUMPS simulations, indicating a first-order phase transition.

\subsection{Spiral spin liquid as continuous phase transition}
Next, we examine the behavior at large $J_H$. In particular, we fix $J_H/J_K = 2$, which is deep inside the spiral phase of the sawtooth spin chain, such that the apex spin correlations are commensurate with period 4. In the main text we consider the path $t_1: 0\rightarrow 1$ while keeping $t_2 = 1$ constant. In the appendix we also show results for a continuation of this path where $t_1=1$ is held constant and $t_2$ decreases to zero. The endpoint of this second segment of the path, i.e. $(t_1,t_2)=(1,0)$, corresponds to the featureless Kondo insulator which adiabatically connects to the trivial strong-coupling state $|\psi_+\rangle$ as $J_H \rightarrow 0$. In the appendix we explicitly check that the states with $t_2=1$ and small but finite $t_1$ are adiabatically connected to this trivial strong-coupling state. We also explicitly checked that there are both singly and multiply degenerate levels in the entanglement spectrum, ruling out non-trivial symmetry-protected topological phases \cite{PhysRevB.81.064439}. 

\subsubsection{Charge gap}
\begin{figure}[t]
    \centering
    \includegraphics[width=\linewidth]{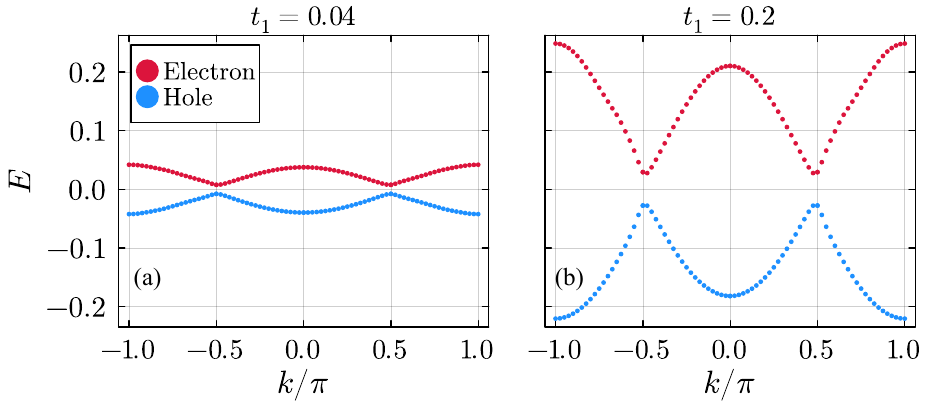}
    \caption{Quasi-particle dispersions for two representative values of $t_1$ at $D = 600$ without enforcing the SU(2) symmetry. We always find a pair of near-degenerate bands for both quasi-electron and -hole cases, corresponding to the spin up and down components.}
    \label{fig:QP}
\end{figure}

In Fig.~\ref{fig:gap}(a) we show electron correlation length $\xi_{e}$ as a function of $t_1/t_2$, for different bond dimensions $D$. We see that $\xi_e$ remains small $(\xi_e \lesssim 4)$ along the entire path, which confirms the insulating nature of the ground state at non-zero $t_1/t_2$. As expected, $\xi_e$ drops to zero at $t_1=0$, where charge correlations are restricted to pairs of neighboring sites. Since the correlation length calculations are performed with fixed electron number, we further confirm the incompressible character of the system by explicitly calculating the quasi-particle and quasi-hole dispersion relations using the MPS quasi-particle ansatz \cite{PhysRevB.85.100408, PhysRevB.88.075133}. The results are plotted in Fig.~\ref{fig:QP} for two representative values of $t_1$, namely $t_1=0.04$ and $t_1=0.2$ \footnote{As mentioned earlier, there are two physical unit cells in each unit cell in our simulation. The quasi-particle dispersions are obtained by unfolding bands in the folded Brillouin zone. We observe weak translation symmetry breaking of order $10^{-3}$ for $t_1 = 0.04$. See also Fig.~\ref{fig:E6v8}. We make an average in plotting Fig.~\ref{fig:QP}(a). However, no averaging is used in extracting the charge gap from the quasi-particle dispersions in Appendix.~\ref{sec:fKI}.}. Note that the dispersion relations are plotted in a way which is familiar from band theory, i.e. by showing the energy of the quasi-particle excitation together with minus the energy of the quasi-hole excitation. Concretely, we plot $\varepsilon_e(k) = E_e(k) - E_0$ and $\varepsilon_h(k) = -(E_h(k)-E_0)$, with $E_e(k)$ and $E_h(k)$ respectively the energy of the many-body state with an additional electron or hole at momentum $k$ on top of the ground state, and $E_0$ is the ground state energy. We can clearly identify a non-zero quasi-particle or charge gap 
\begin{eqnarray}\label{defDelqp}
\Delta_{qp} & = & \underset{k,k'}{\text{min}}\left(E_e(k) + E_h(k') - 2E_0 \right) \\
 & = & \underset{k}{\text{min}}\,\varepsilon_e(k) - \underset{k}{\text{max}}\,\varepsilon_h(k)  \,,\nonumber
\end{eqnarray}
which confirms that the system is indeed incompressible at $t_1=0.04$ and $t_1=0.2$. 

\begin{figure}
    \centering
    \includegraphics[width=\linewidth]{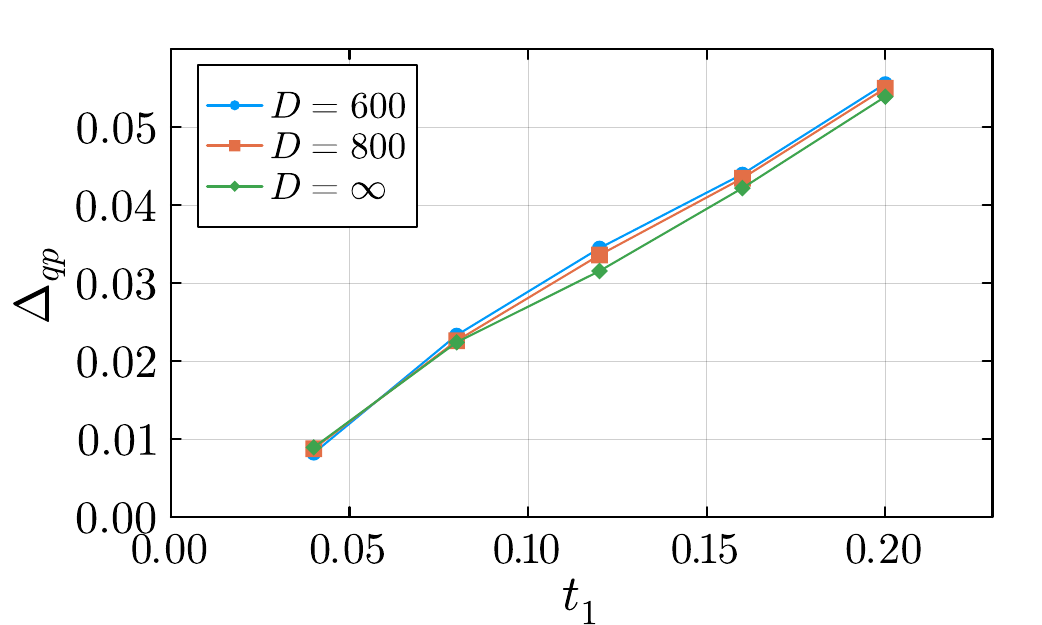}
    \caption{Charge gap $\Delta_{qp}$ as defined in Eq.~\eqref{defDelqp} as a function of $t_1$ at $J_H=1.2$.}
    \label{fig:qpgapvt1}
\end{figure}

In Fig.~\ref{fig:qpgapvt1} we plot $\Delta_{qp}$ as a function of $t_1$. We see that the charge gap quickly decreases as $t_1\rightarrow 0$, and approaches zero. If $\Delta_{qp}$ were to vanish at finite $t_1$, this would give rise to a diverging electron correlation length $\xi_e$. But we find that $\xi_e$ remains small for all $t_1$, and are therefore lead to the conclusion that the system becomes compressible exactly at $t_1=0$.

\subsubsection{Spin correlations}
Next, we consider the spin correlations. In Fig.~\ref{fig:gap}(b) we see that the spin correlation length $\xi_m$, interpolated to infinite bond dimension, evolves non-monotonically and displays a pronounced dip near $t_1/t_2 \sim 0.02$. To rule out the possibility that this dip in $\xi_m$ is a kink induced by a first order phase transition we compute the ground state fidelity density in Fig.~\ref{fig:gap}(c), which for two states $|\psi_1\rangle$ and $|\psi_2\rangle$ on a system with $L$ unit cells is defined as 
\begin{equation}
f = \lim_{L\rightarrow \infty} |\langle \psi_1|\psi_2\rangle|^{1/L}\label{deff}
\end{equation}
This fidelity density is thus a measure for the overlap per unit cell of the two states, and is equal to 1 iff the two states are identical up to a phase. For two translationally-invariant MPS with matrices $A^i_1$ and $A^i_2$ the fidelity density is given by the magnitude of the leading eigenvalue of the mixed transfer matrix $\mathcal{T}_{12} = \sum_i \bar{A}^i_1\otimes A^i_2$, which is how we calculate $f$ in practice. In Fig.~\ref{fig:gap}(c) we show $f$ obtained from the MPS at $t_1$ and $t_1+\delta t_1$. A first-order phase transition would manifest as a sudden dramatic change in the ground state, and hence cause the fidelity obtained from the MPS at two different sides of the transition being markedly smaller than the fidelities obtained from pairs of states in the same phase. Fig.~\ref{fig:gap}(c) shows no outlier in the fidelities, which rules out a first-order phase transition. We do see that on the left of the dip in $\xi_m$, i.e. for $t_1/t_2\lesssim 0.02$, the fidelities become smaller, indicating that the ground state changes more rapidly with $t_1/t_2$ in this parameter range.

\begin{figure*}[t]
    \centering
    \includegraphics[width=\linewidth]{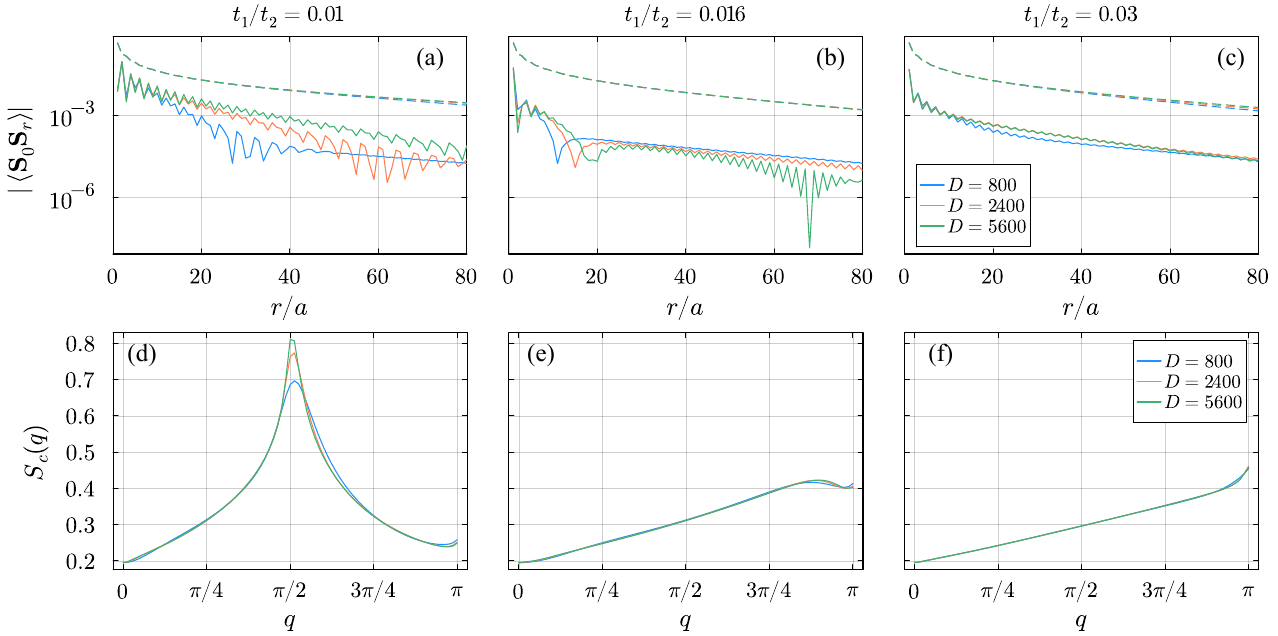}
    \caption{Real-space spin correlation functions (a-c) of the $c$-electrons (solid) and $f$-moments (dashed), and corresponding static structure factors (d-f) for the $c$-electrons, obtained at three different values of $t_1$ and with three different bond dimensions. The other parameters used for these simulations were $J_H=1.2$, $J_K=0.6$, and $t_2=1$.}
    \label{fig:insulators}
\end{figure*}

To understand the physical origin of the dip in $\xi_m$, we examine the real-space spin correlation functions and the static spin structure factors (the Fourier transform of the spin correlation functions) in Fig.~\ref{fig:insulators}. The real-space spin correlations of the $c$- and $f$-electrons display distinct features, with the spin correlations in the $c$ chain being much smaller in magnitude and having a shorter correlation length. In Fig.~\ref{fig:insulators}(d-f), we plot the static spin structure factor for the $c$ electrons. Notably, when $t_1/t_2$ is small, the spin correlations in the $c$ electrons resemble those of the apex spins in the sawtooth spin chain, with a peak in the spin structure factor at $k=\pi/2$. In contrast, when $t_1/t_2\gtrsim 0.01$, this peak starts to continuously shift away from $\pi/2$ until it reaches $k=\pi$ for $t_1/t_2 \gtrsim 0.02$. For these larger values of $t_1/t_2$, both the $c$-electrons and the $f$-moments display anti-ferromagnetic correlations. We thus see that the dip in $\xi_m$ coincides with the point at which the peak in the structure factor of the $c$-electrons starts to move away from $\pi$. One possibility is therefore that $t_1/t_2\sim 0.02$ is a disorder point~\cite{denNijs}, similar to the AKLT point in the spin-$1$ chain with biquadratic interactions~\cite{Schollwock1996,Herviou2024}, and the Majumdar-Ghosh point in the $J_1-J_2$ spin-$1/2$ chain~\cite{White1996}, which also display a kink in the correlation length at the onset of incommensurate short-range correlations. However, the $t_1/t_2\sim 0.02$ point in our model differs from these previously studied disorder points in some potentially important ways. First, the minimal correlation length is $\sim 15$ lattice sites and the MPS representation of the corresponding ground state requires a high bond dimension, whereas the AKLT and Majumdar-Ghosh states have a correlation length of order lattice spacing and admit exact low-bond dimension MPS representations. Also, in our Kondo lattice model the $c$-electron correlations become incommensurate, but the dip we observe is in the correlation length of the $f$-moments. The $c$-electrons and $f$-moments are coupled of course, but the correlations of the latter remain commensurate and fixed at $k=\pi$ upon crossing the putative disorder point $t_1/t_2\sim 0.02$.

Importantly, the spin correlation length $\xi_m$ extrapolated to infinite bond dimension remains finite for all non-zero $t_1/t_2$ used in our simulations. Together with the fact that the Kondo lattice system is incompressible for non-zero $t_1/t_2$, and does not break the reflection or translation symmetry~\footnote{In our simulations we use a unit cell which consists of two physical unit cells, so a spontaneous doubling of the unit cell would be detectable. Larger periodicities have not been explicitly investigated, but from the fact that the MPS remains injective we can safely rule out any form of spontaneous translation symmetry breaking.}, this shows that the spiral spin liquid of the sawtooth spin chain represents a critical point between two classes of featureless Kondo insulators adiabatically connected to either $|\psi_+\rangle$ ($t_1>0$, as explicitly studied in our numerics) or $|\psi_-\rangle$ ($t_1<0$). 

Lastly, let us come back to the fact that also the charge gap closes at $t_1=0$. In Ref.~\cite{sawtooth} it was shown that the commensurate spiral spin liquid has a flat (i.e. non-dispersing) zero-energy spectral weight near $k=\pm \pi/2$ associated with fluctuations of the apex spins. Based on this property, we suggest the following physical picture for the vanishing charge gap. When $t_1=0$, the electron or hole doped into the system nucleates a localized and dispersionless spinon excitation of the spiral spin liquid. This spinon screens the spin of the doped electron or hole, leaving behind a localized charge which costs zero energy. We emphasize that this picture only applies when we add a single electron or hole into the system, not when we add multiple or even a finite density of quasi-particles. To see this, consider the extreme case where we dope a hole in every unit cell, such that we completely remove all the apex spins. In this way we obtain the pure Heisenberg model for the basal spins, which has a ground state energy of $E_{\text{Heis}} \sim -0.4431\, J_H $ per unit cell. On the other hand, for the spiral spin liquid with an apical-basal spin exchange coupling of $0.25J_H$ we find a ground state energy of $E_{\text{spiral}} \sim -0.4534\, J_H$ per unit cell (from $D=5600$ MPS simulations). This shows that removing all the apex spins does indeed cost a finite energy density.

\begin{figure}
    \centering 
    \includegraphics[width=\linewidth]{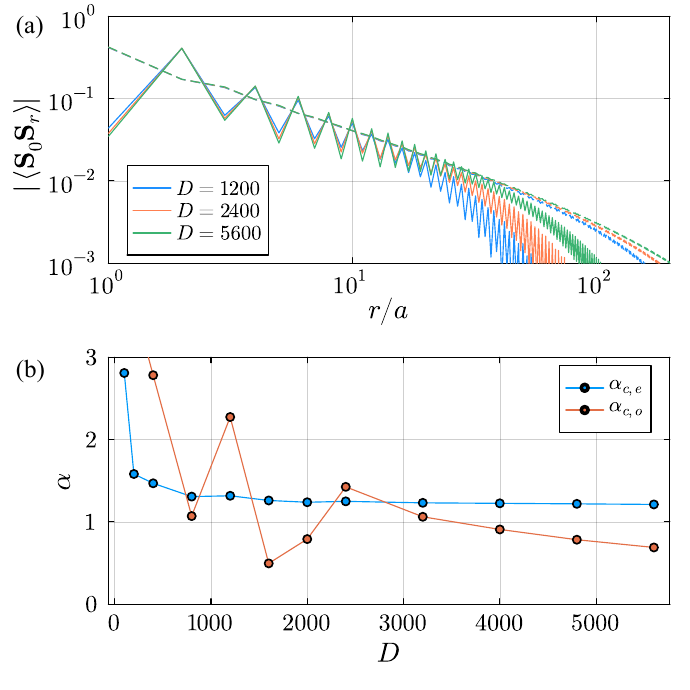}
    \caption{(a) Magnitude of spin correlations in the sawtooth spin chain. The $c$- and $f$-chain show distinct spin correlations: the non-oscillating correlations (dashed) come from the $f$-chain, with asymptotic Heisenberg behavior $1/r^{\alpha_f}$ with $\alpha_f = 0.995 \approx 1$. The $c$-chain correlations (solid) show an even-odd effect, with the odd distances being less converged in bond dimension. (b) The spin correlation decay exponents $\alpha_{c,e}$ and $\alpha_{c,o}$ for even and odd separation in the $c$-chain as a function of bond dimension (see Eq.~\eqref{spinc}).}
    \label{fig:QDS}
\end{figure}

\section{Nature of the spiral spin liquid} 
Given that we have identified the gapless spiral spin liquid of the sawtooth spin chain as a critical point between two fragile Kondo insulators, let us revisit its properties in more detail. To obtain our results for the sawtooth spin chain presented below we have always imposed explicit SU(2) symmetry on the MPS. Following Ref.~\cite{sawtooth}, we separately analyze the magnetic correlations in the $c$- and $f$-chains, as shown in Fig.~\ref{fig:QDS}(a). Both chains exhibit power-law decay at short distances, transitioning to exponential decay at longer distances as a result of the finite bond dimensions used in our simulations. Assuming that the $f$-spin correlations follow a simple asymptotic form $\<{\mathbf{S}_0 \cdot \mathbf{S}_r}\> \sim \cos(\pi r)/r^{\alpha_f}$, we fit the power-law regime to extract the exponent $\alpha_f$. At the largest bond dimension accessible in our simulations, the $f$-chain behaves like a Heisenberg chain, with the numerically obtained $\alpha_f = 0.995$ very close to the exact value $\alpha_{\text{heis}}=1$. The correlations in the $c$-chain, however, display following more complex behavior:
\begin{equation}
\<{\mathbf{S}_0 \cdot \mathbf{S}_r}\> \sim A\frac{\cos(\pi r/2)}{r^{\alpha_{c,e}}} + B \frac{\sin(\pi r/2)}{r^{\alpha_{c,o}}}\,,\label{spinc}
\end{equation}
where $A$ and $B$ are two constants. We find that the two exponents $\alpha_{c,e}$ and $\alpha_{c,o}$ for even and odd $r$ are different, and both deviate significantly from the Heisenberg value $\alpha=1$. In Fig.~\ref{fig:QDS}(b) we show the numerically obtained exponents $\alpha_{c,e}$ and $\alpha_{c,o}$ as a function of bond dimension. The exponent corresponding to even $r$ converges much faster with $D$, and remains more or less unchanged for $D>2000$. The exponent for odd $r$, however, converges much slower. Even at $D=5600$ $\alpha_{c,o}$ is not yet fully converged, but does start to vary more slowly. Importantly, at $D=5600$ it is clear that $\alpha_{c,o}$ continues to drift away from the converged value of $\alpha_{c,e}$, such that we can be quite confident that they are really distinct.

To further characterize the gapless spiral spin liquid we computed the entanglement entropy for different bond dimensions and compared it to the universal CFT relation $S = \frac{c}{6}\log\xi_m + \text{const}$. The quality of a fit to this relation is much worse than in the SU(2)$_1$ phase of the sawtooth spin chain, and produces an unreasonably large value of $c\sim 4.4$ (see supplementary material). Additionally, we performed Exact Diagonalization (ED) calculations for system sizes up to 17 unit cells (34 spins) under both open and periodic boundary conditions to extract the finite-size gap $\Delta_L$. The gap $\Delta_L$ shows strong deviations from $L^{-z}$ scaling for any fixed dynamical exponent $z$, as the fitted $z$ value depends sensitively on both boundary conditions and system size, as can be seen in the supplementary material. Because of these results we expect that the spiral spin liquid is not described by a CFT. We cannot reliably extract the dynamical exponent because not only does the $\Delta_L \sim L^{-z}$ scaling fail, also the MPS quasi-particle ansatz to extract the dispersion relation of low-energy excitations fails. It thus remains possible that $z=\infty$, as suggested by the regions of flat zero-frequency spectral weight in the dynamic spin structure factor calculated in Ref.~\cite{sawtooth}.

\section{Discussion} 
We have shown that the multi-orbital Kondo lattice model in Eq.~\eqref{defH} hosts a VBS phase, an extended SU(2)$_1$ spin liquid region, and two distinct fragile Kondo insulators. We furthermore argued that the fragile Kondo insulators are separated by a critical line which corresponds to the gapless spiral spin liquid of the sawtooth spin chain. To the best of our knowledge, this constitutes the first example of a Kondo insulator exhibiting the non-trivial spin physics of VBS states and spin liquids typically associated with Mott insulators. But perhaps the most surprising feature in the phase diagram is the direct continuous transition between the two fragile Kondo insulators. We are aware of only one other example of a continuous phase transition between two trivial featureless phases solely distinct by symmetry quantum numbers. In particular, Refs.~\cite{Tsukano1998,SPt} found such a transition in a spin-1 chain with broken spin rotation symmetry. In that model, the symmetry which distinguishes the two trivial phases is a combination of site-centered reflection and a $\pi$ spin rotation around the $z$-axis~\cite{SPt}. Our Kondo lattice model differs from this spin-1 chain in several ways. First, our model is SU(2) symmetric and the critical point has a LSM obstruction due to this SU(2) symmetry (and reflection). Secondly, charge fluctuations are crucial in the Kondo lattice model. To see this, assume that the $c$-electrons are also Mott-localized. In this case adding a small $t_1$ hopping introduces a nearest-neighbor super-exchange coupling for the $c$-electron spins. But the LSM obstruction remains, and hence the system either has to remain gapless or break reflection symmetry -- a transition into a featureless state is impossible. And lastly, for the spin-1 chain the nature of the critical point is well-understood: it corresponds to a Luttinger liquid~\cite{SPt}. On the other hand, for the Kondo lattice model we have argued above that the spiral spin liquid is quite exotic and does not correspond to a CFT in the infrared. The spiral spin liquid is similar to the concept of the ``floating phase'' \cite{bak1982commensurate} in that the incommensurate correlation wave vector changes continuously within the phase. However, known examples of the floating phase, including Rydberg atom ladders \cite{floatingRydberg, zhang2024probingquantumfloatingphases}, frustrated spin-5/2 chains \cite{floating5halves}, interacting Majorana chains \cite{Majoranachain}, and quantum Ashkin-Teller chains \cite{Ashkin-Teller}, are still described by the Luttinger liquid theory.

Adding a small on-site Hubbard repulsion for the $c$-electrons will cause the latter to be Mott-localized for very small $t_1$ (but keeping $t_2=1$). As a result, the spiral spin liquid will extend into a narrow 2D region in the $(t_1,J_H)$ plane, showing up as an intermediate phase in between the two fragile Kondo insulators. The continuous transition we find in this work is therefore really a multi-critical point where the two fragile Kondo insulators and the Mott-localized spiral spin liquid meet.

Finally, let us also mention one drawback of the Kondo lattice model in Eq. \eqref{defH}: it is not easily obtained from an Anderson model. In the appendix we show that it is possible to obtain Eq. \eqref{defH} as a low-energy description of an Anderson model, but the construction is somewhat contrived. Nevertheless, we believe that the results presented here provide a motivation to look for other variations of Kondo lattice models which also host non-trivial spin physics, but which have a higher chance of being realized in Nature. In some respects our work is similar in spirit to the recent papers~\cite{Glittum2024} and~\cite{Cai2024}, which find spin liquids stabilized by kinetic energy frustration and electron-phonon coupling. So similar to our findings in the multi-orbital Kondo model, these works also suggest to start looking for spin liquids outside of the usual arena of Mott insulators with geometric frustration. Finally, an interesting direction for future work is to study what happens upon doping the $c$-electrons away from quarter-filling, especially given the rich physics already present in the conventional 1D doped Kondo-Heisenberg model \cite{Fujimoto1994,Moukouri1996,Sikkema1997,Xavier2008,Berg2010,Eidelstein2011,Yang2025}.

{\it Acknowledgments} --- This research was supported by the European Research Council under the European Union Horizon 2020 Research and Innovation Programme via Grant Agreement No. 101076597-SIESS (N.B. and N.H.).

\bibliography{ref}

\clearpage
\newpage
\onecolumngrid

\appendix

\section{Classical considerations for the sawtooth spin chain}
We consider a zigzag Heisenberg spin ladder, similar to Ref.~\cite{White1996}:
\begin{align}
    H_2 = J_{BB}\sum_{\<{i,j}\>}\bm{S}_{i, B}\cdot \bm{S}_{j, B} + J_{AA}\sum_{\<{i,j}\>}\bm{S}_{i, A}\cdot \bm{S}_{i, A} + J_{AB} \sum_{\<{i,j}\>}\(\bm{S}_{i, B}\cdot \bm{S}_{i, A} + \bm{S}_{i, A}\cdot \bm{S}_{i+1, B}\), 
\end{align}
where $i$ labels the unit cell and $A/B$ should be considered sublattices. In the following we will take $J_{BB} = 1$. $J_{AB}$ corresponds to $ J_K$ in the main text, and $J_{AA}=0$ gives the sawtooth limit. We first apply the Luttinger-Tisza-Lyons-Kaplan method \cite{luttinger-tisza, lyons-kaplan} to study the possible ground state phases. Taking the spins to be classical objects with unit length, the generalized weak constraint is
\begin{align}
    \sum_i\sum_{\nu} \alpha_{\nu}^2 \bm{S}^2_{i, \nu}= N\sum_{\nu} \alpha_{\nu}^2,
\end{align}
where $\nu$ is the sublattice label. $\alpha_{\nu}$ are real numbers introduced to further relax the weak constraint such that it's possible to rescale the two sublattices so that they are equivalent, which is the case where the original Luttinger-Tisza method works.  Using translation symmetry, we rewrite $\bm{S}_{i,\nu} = \sum_k \bm{Q}_{k,\nu}\exp(ikr_{i,\nu})$. In $k$-space, we find that the energy $E/N = \sum_k J_{\mu,\nu}(k)\bm{Q}_{k,\mu}^*\bm{Q}_{k,\nu}$ is determined by
\begin{align}
    \mat{J}(k) = \begin{pmatrix}
    J_{AA}\cos k & J_{AB}\cos (k/2) \\
    J_{AB}\cos (k/2) & \cos k
    \end{pmatrix},
\end{align}
where the lattice constant is omitted, with the constraint $\sum_k \sum_{\nu} \alpha_{\nu}^2|\bm{Q}_{k,\nu}|^2 = \sum_{\nu}\alpha_{\nu}^2$. Introducing Lagrange multipliers for the constraint and minimizing energy with the constraint yields
\begin{align}
    \mathcal{J}_{\mu\nu}(k)\bm{P}_{k,\nu} = \epsilon \bm{P}_{k,\mu},
\end{align}
where we used $\mathcal{J}_{\mu\nu} = \alpha^{-1}_{\mu}\alpha^{-1}_{\nu}J_{\mu\nu}$, and $\bm{P}_{k,\nu} = \alpha_{\nu}\bm{Q}_{k,\nu}$. In this notation, the constraint is $\sum_k \sum_{\nu} |\bm{P}_{k,\nu}|^2 = \sum_{\nu}\alpha_{\nu}^2$. Let $k_0$ be the momentum that minimizes $\epsilon(k, \alpha)$. We then choose to parameterize $\bm{P}_{k,\nu}$ to be $\bm{c}\psi_{\nu}$ at $\pm k_0$, where $(\psi_A,\psi_B)^T$ is the eigenvector, and zero at all other momenta. Choosing the spin to sit in the $x-y$ plane, we can write
\begin{align}
    \bm{S}_{i,\nu} = |\psi_\nu|/\alpha_{\nu}\[ \hat{x}\sin\(k_0r_{i}+\phi_{\nu}\) + \hat{y}\cos\(k_0r_{i}+\phi_{\nu}\)\].\label{eq:k0}
\end{align}
It should be clear now that $\alpha_{\nu}$ allows for $|\psi_A| \neq |\psi_B|$, and is fixed by the strong constraint $\bm{S}_{i,\nu}^2 = 1$. When the two sublattices are identical, $|\psi_A| = |\psi_B| = 1/\sqrt{2}$.

For $J_{AA}=0$, we explicitly write out the eigenequation for
\begin{align}
    \mat{\mathcal{J}}(k) = \begin{pmatrix}
    0 & J_{AB}\alpha_A^{-1}\alpha_B^{-1}\cos (k/2) \\
    J_{AB}\alpha_A^{-1}\alpha_B^{-1}\cos (k/2) & \alpha_B^{-2}\cos k
    \end{pmatrix}. 
\end{align}
and its lower energy eigenvector $(\psi_A,\psi_B)^T = \(-\sin(\theta_k/2), \cos(\theta_k/2)\)^T$, with $\cos\theta_k = -h_0/\epsilon_k$. Here $h_0 = \alpha_B^{-2}\cos k/2$.
Without loss of generality, we choose $\alpha_A = |\psi_A|,\ \alpha_B = |\psi_B|$ such that the strong constraint can be satisfied. In general it's difficult to minimize the energy now since the parameters $\alpha_{A/B}$ enter in both the matrix $\mat{\mathcal{J}}(k)$ and its eigenvectors. However, we know the energy for a spiral ansatz Eq.~\eqref{eq:k0} is 
\begin{align}
    E(k_0) &= \min_{\theta}\{\cos k_0 + J_{AB}[\cos\theta + \cos(k_0-\theta)]\} \nonumber\\
    &= \cos k_0 + 2J_{AB}\cos\frac{k_0}{2}.
\end{align}
where $\theta = \phi_A-\phi_B$ is chosen to be $k_0/2$ from the first line to the second line. Hence the minimum $k_0$ is given by $\cos (k_0/2) = -J_{AB}/2$, i.e. an minimum exist when $|J_{AB}|<2$. With this, we find the minimum classical energy $E_0 = 2(-J_{AB}/2)^2 -1 -J_{AB}^2 = -1 - J_{AB}^2/2$. In the specific case of $J_{AB} \ll 1$, this means $k_0 \approx \pi + J_{AB}$. In other words, both sublattices are incommensurate spirals with almost $\pi/2$ phase difference.  

Notice that the quantum ground state obtained in the main text has a different classical configuration, which sets the $B$ sites to form the N\'eel state and the $A$ sites form a spiral order with angle $\phi$. The classical energy density is then $E = -1 + J_{AB} \( \cos( \theta ) + \cos( \pi-\theta ) \) + J_{AA}\cos \phi = -1 + J_{AA}\cos \phi$, where $\theta$ is again the relative spin angle between $A$ and $B$ sites. We see disordering in the $A$ sites at the classical level when $J_{AA}=0$.

\section{Central charge of $SU(2)_1$ spin liquid}
\begin{figure}
    \centering
    \includegraphics[width=0.5\linewidth]{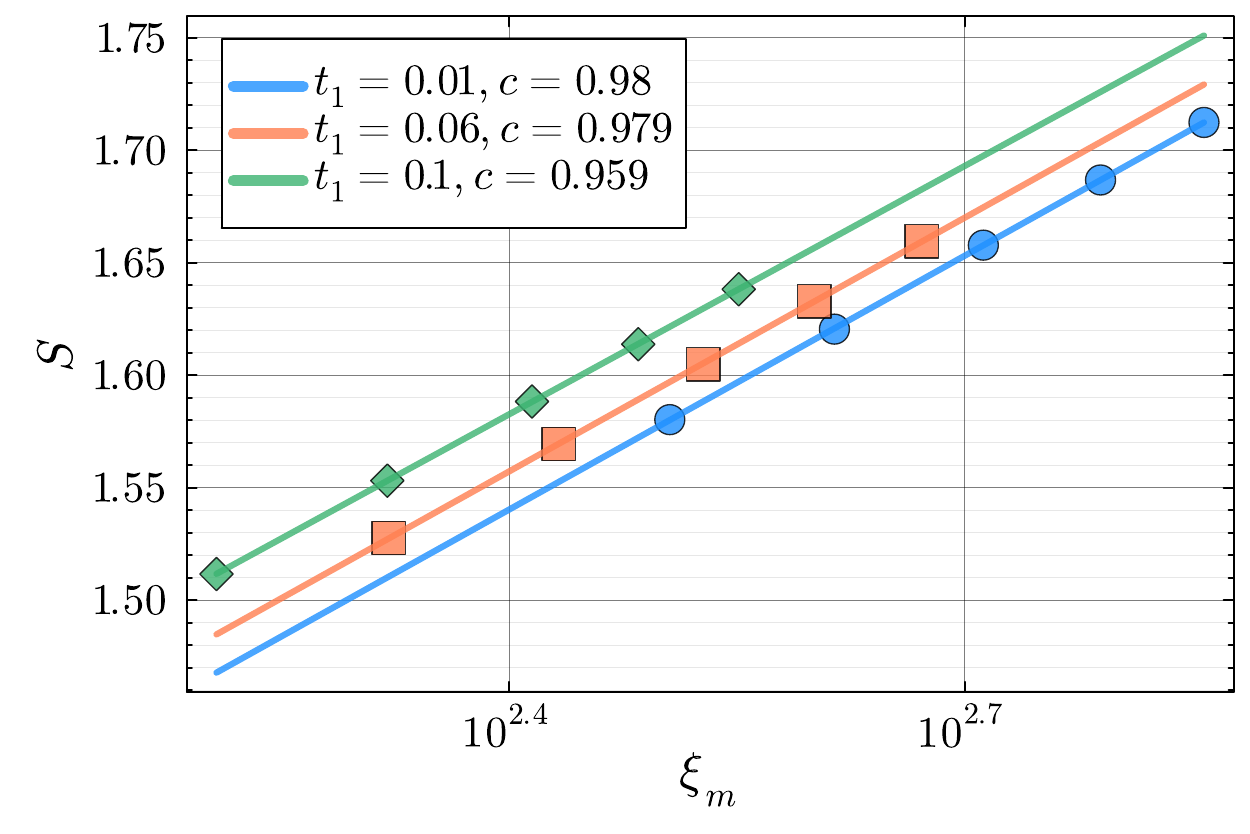}
    \caption{Entanglement entropy $S$ scaling with correlation length $\xi_m$ for a few $t_1$ values at $J_H = 0.05$, i.e. in the $SU(2)_1$ spin liquid phase. Data points come from $D = [800, 1000, 1200, 1400, 1600]$.}
    \label{fig:SU2_1}
\end{figure}
In Fig.~\ref{fig:SU2_1} we show that for small $J_H$ the universal CFT relation $S = \frac{c}{6}\log\xi_m + \text{const}$ between the entanglement entropy $S$ of the half-infinite chain and the bond-dimension-induced correlation length $\xi_m$ is satisfied over a finite range of $t_1$, with $c\approx 1$. For this range of hoppings the system is therefore in the same gapless phase as the Heisenberg chain. We note that for this phase, the entanglement entropy is much smaller than in the spiral spin liquid phase (see Fig.~\ref{fig:Svxi}) for the same bond dimension. In particular, for $t_1 = 0.01$, a bond dimension of $1600$ already yields a correlation length of $720$, in stark contrast to the spiral spin liquid phase.

\section{More details on the featureless Kondo insulators at large $J_H$} \label{sec:fKI}
In Fig.~\ref{fig:phases} we show a color plot of the reflection symmetry-breaking order parameter $O_{\mathcal{R}}$ up to $J_H = 0.6$. Here we now plot the numerical values of $O_{\mathcal{R}}$ at $J_H = 1.2$, which is the value of $J_H$ we focus on in most of the paper. At $D = 300$ we see reflection symmetry breaking which already very weak since $O_{\mathcal{R}} \sim \mathcal{O}(10^{-5})$. We therefore assume this to be a finite-$D$ effect and consider this phase to be reflection symmetric.
\begin{figure}
    \centering
    \includegraphics[width=0.5\linewidth]{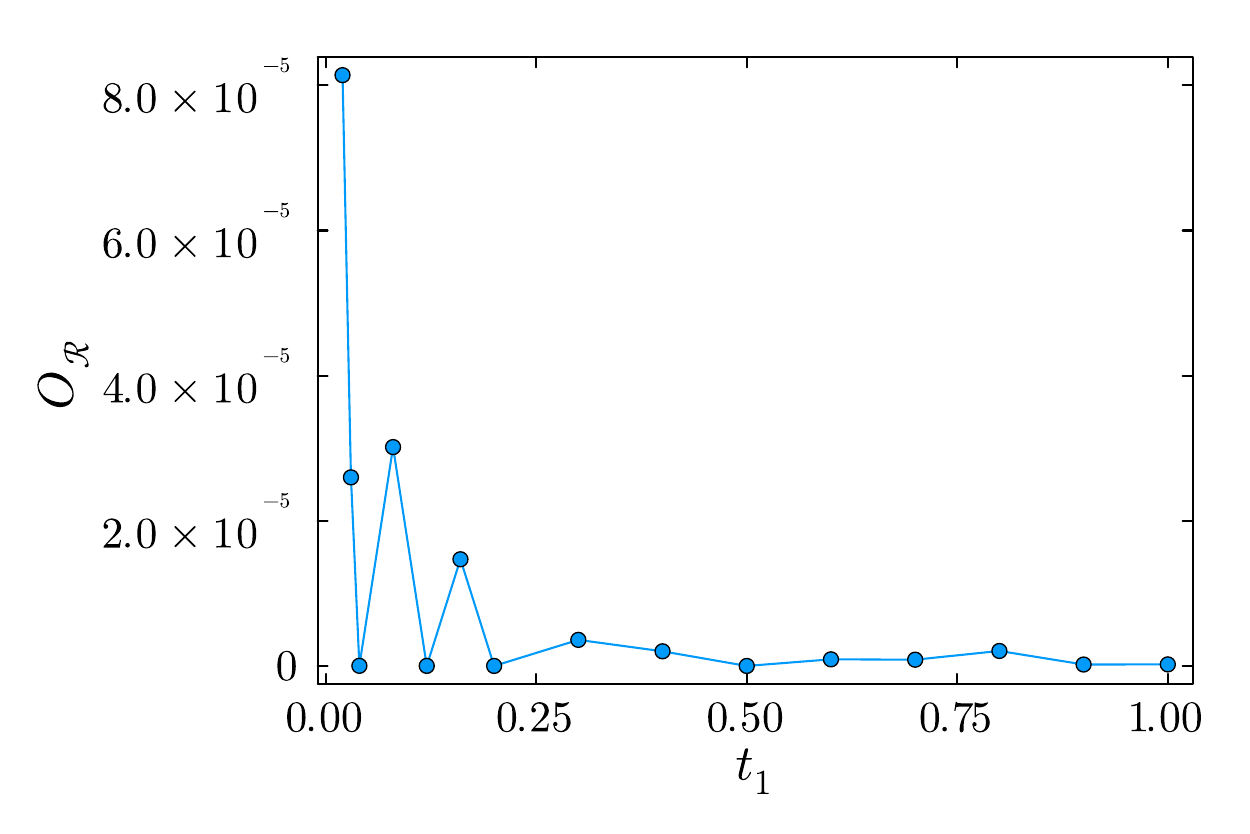}
    \caption{Numerical values of the reflection symmetry-breaking order parameter $O_{\mathcal{R}}$ for $J_H = 1.2$ and $t_2 = 1.0$ at $D = 300$.}
    \label{fig:OR300}
\end{figure}

\begin{figure}
    \centering
    \includegraphics[width=\linewidth]{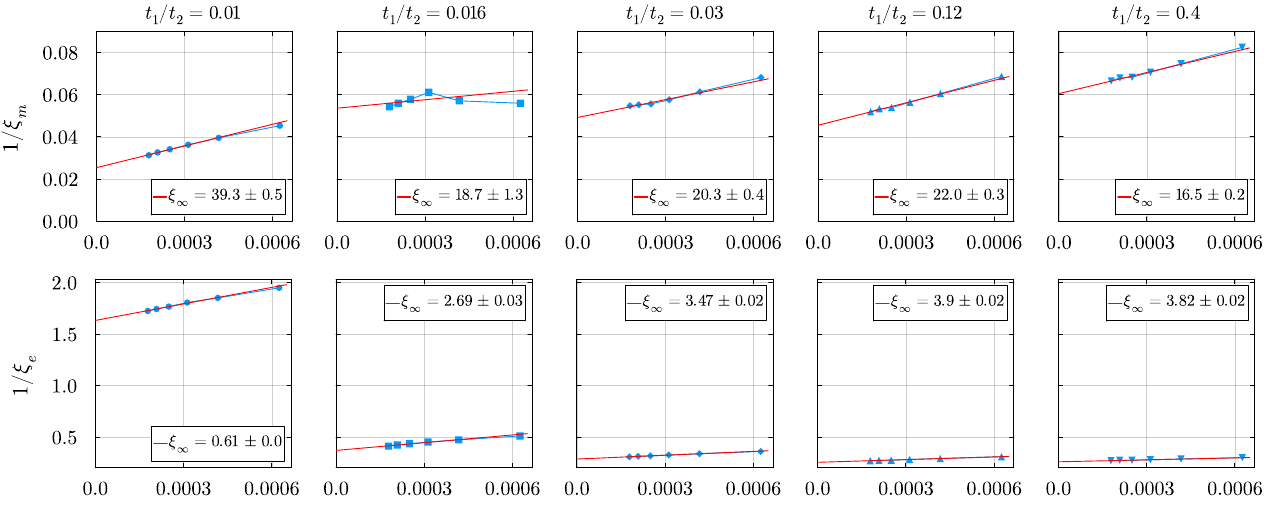}
    \caption{Examples of the extrapolation of magnetic and electron correlation lengths to infinite $D$.}
    \label{fig:fitxi}
\end{figure}

In Fig.~\ref{fig:gap}, we show that the featureless Kondo insulators are indeed featureless, and hence gapped in both spin and charge sectors by extrapolating the correlation lengths to infinite $D$. Some examples of these extrapolations are shown in Fig.~\ref{fig:fitxi}. Data points are obtained with $D = [1600, 2400, 3200, 4000, 4800, 5600]$, but fittings are performed excluding $D = 1600$. 

\begin{figure}
    \centering
    \includegraphics[width=0.7\linewidth]{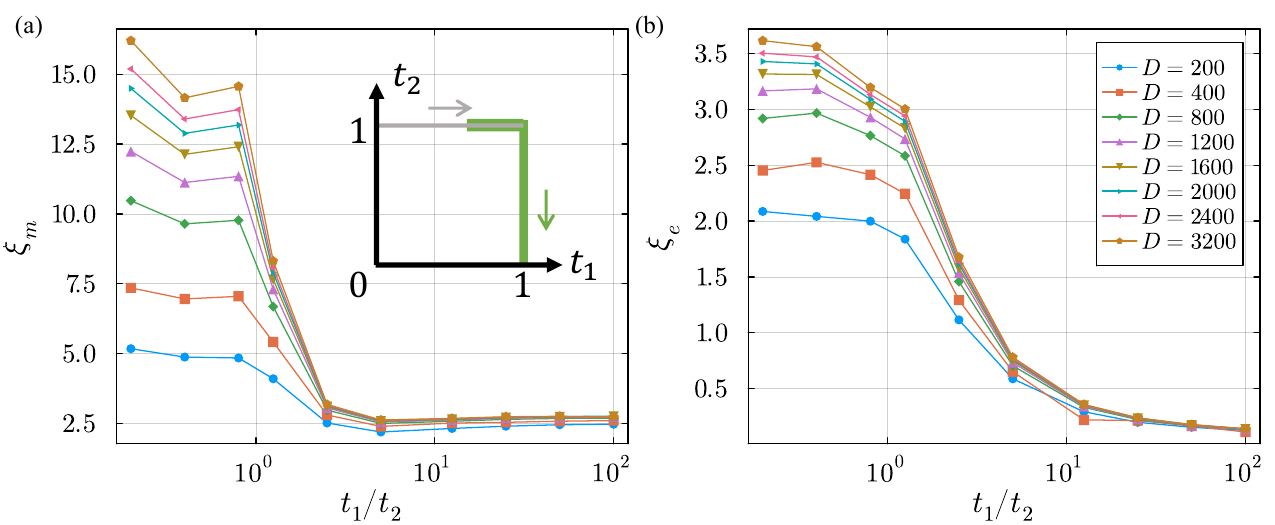}
    \caption{Spin (a) and electron (b) correlation lengths $\xi_{m/e}$ as a function of $t_1/t_2$ for different bond dimensions $D$ along the green path. The path used in Fig.~\ref{fig:gap} is shown in gray.}
    \label{fig:larget1}
\end{figure}

To make it clear that the featureless Kondo insulator is indeed adiabatically connected to the product state limit ($t_2 = 0$), we show an additional path (green in inset) where $t_1 = 1$ and $t_2 \rightarrow 0$ in Fig.~\ref{fig:larget1}. (The first three points are on the path shown in Fig.~\ref{fig:gap}.) We see clearly that both $\xi_m$ and $\xi_e$ decay quickly as $t_1/t_2$ increases. They also saturate faster as $D$ increases.

\begin{figure}
    \centering
    \includegraphics[width=\linewidth]{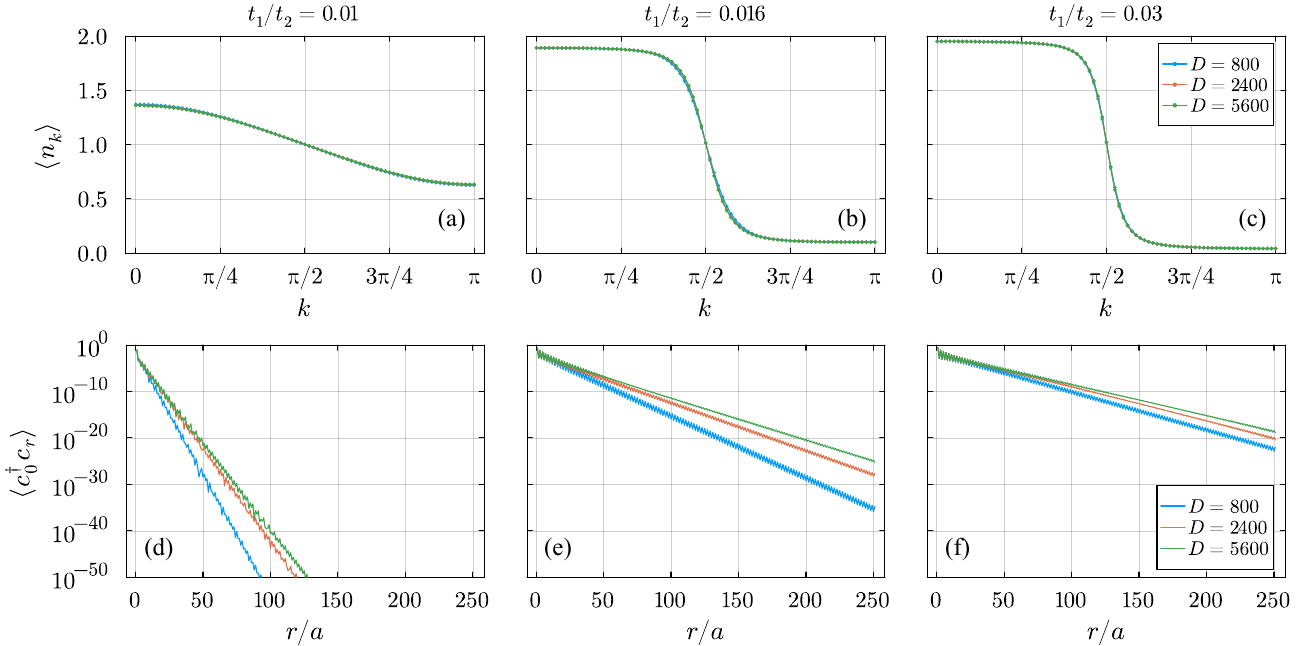}
    \caption{The $c$-electron momentum distribution function $\langle n_k \rangle$, and two-point function $\langle c_0^{\dagger}c_r \rangle$.}
    \label{fig:nk}
\end{figure}

In addition to the correlation lengths, we also examine the $c$-electron momentum distribution function $\langle n_k \rangle = \langle c_{k, A}^{\dagger}c_{k, A} \rangle + \langle c_{k, B}^{\dagger}c_{k, B} \rangle$ in Fig.~\ref{fig:nk}(a-c) for a few values of $t_1$. $\langle n_k\rangle$ is a smooth function of momentum, which is compatible with an insulating state at any finite $t_1$. We also plot in Fig.~\ref{fig:nk}(d-f) the two-point function $\langle c_0^{\dagger}c_r \rangle \equiv (\langle c_{0, A}^{\dagger}c_{r, A} \rangle + \langle c_{0, B}^{\dagger}c_{r, B} \rangle)/2 = \langle c_{0, A}^{\dagger}c_{r, A} \rangle = \langle c_{0, B}^{\dagger}c_{r, B} \rangle$, where we used the lack of symmetry breaking in reflection for the last two equalities. An exponential decay is observed for all $t_1$ values. From the slopes in Fig.~\ref{fig:nk} (d-f) we see an increase in correlation length $\xi_e$ as $t_1$ increases, as also shown in Fig.~\ref{fig:gap}(b) in the main text.

\begin{figure}
    \centering
    \includegraphics[width=\linewidth]{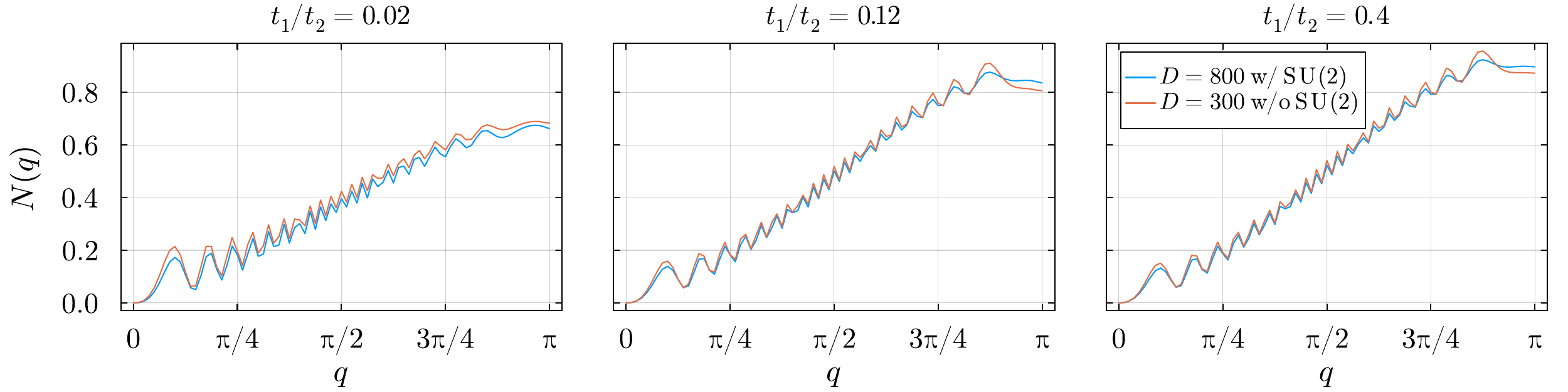}
    \caption{Charge structure factor $N(q)$ computed with and without SU(2) symmetry.}
    \label{fig:csf}
\end{figure}

We next consider the charge structure factor 
\begin{align}
    N(q)  = \frac{1}{L}\sum_{i,j}e^{iq(r_i-r_j)}\(\<{n_in_j}\> - \<{n_i}\>\<{n_j}\>\), \label{eq:nq}
\end{align}
where $i, j$ labels unit cells and $n_i = n_{i,A} + n_{i,B}$ is the electron number operator for the $i$th unit cell. Another useful way to write Eq.~\eqref{eq:nq} is through the density fluctuation $\delta n_q = 1/\sqrt{L}\sum_j e^{iqr_j}(n_j - \<{n_j}\>)$:
\begin{align}
    N(q)  = \<{\psi_0}| \delta n_{-q}\delta n_{q} |{\psi_0}\>,
\end{align}
where we explicitly write out the expectation evaluated on the ground state function $|{\psi_0}\>$. A charge excitation $|{\psi}\> = \delta n_q|{\psi_0}\>$ then has energy
\begin{align}
    E_q = \frac{\<{\psi|H|\psi}\>}{\<{\psi|\psi}\>}.
\end{align}
Note that at $q = 0$, $\delta n_{q=0}|{\psi_0}\> = 0$. In the small-$q$ limit, we have
\begin{align}
    \lim_{q\rightarrow 0}(E_q-E_0) = \frac{q^2}{2N(q)} \<{\psi_0}| \int\frac{dk}{2\pi}\d_k^2\epsilon_k c_k^{\dagger}c_k |{\psi_0}\>.
\end{align}
Hence $N(q) \sim q^2$ at small $q$ is a necessary condition for a charge gap (near $q=0$) \cite{Capello2005}. In Fig.~\ref{fig:csf}, we see that $N(q)$ vanishes faster than $q$ for all values of $t_1$ as $q\rightarrow 0$, which provides another consistency check that the system is indeed insulating for all non-zero $t_1$.

\subsection{Simulations with and without SU(2) symmetry}
\begin{figure}
    \centering
    \includegraphics[width=\linewidth]{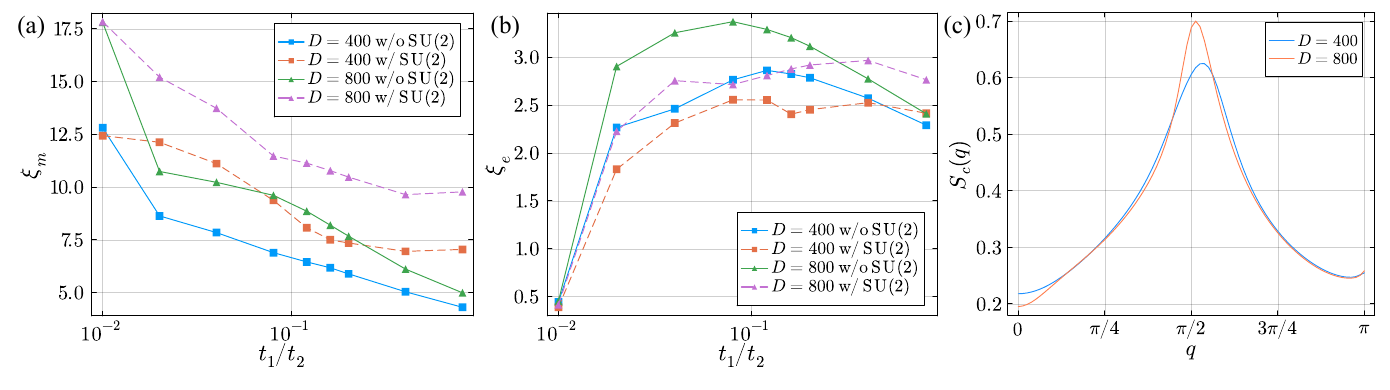}
    \caption{Comparison of correlation length $\xi_m$ in (a) and $\xi_e$ in (b) between simulations with and without SU(2) symmetry. (c) shows the $c$-electron spin structure factor at $t_1 = 0.01$ and $J_H = 1.2$.}
    \label{fig:NSU2}
\end{figure}

In Fig.~\ref{fig:csf}, we also show that the charge structure factors obtained in simulations with and without SU(2) symmetry enforced are similar. In Fig.~\ref{fig:NSU2} we show a few other characteristics that non-SU(2)-symmetric and SU(2)-symmetric MPS share. This is a non-trivial check since at low bond dimensions the MPSs are approximations to the true ground state, and the symmetric and non-symmetric states generically have a different amount of entanglement, and the symmetric states are sensitive to LSM constraints whereas the non-symmetric ones are not.

We do find differences in the quasi-particle dispersions at $J_H = 1.2$ starting from symmetric or non-symmetric ground state MPS. For SU(2) symmetric MPS, we find that $\Delta_{qp}$ sometimes becomes negative, which indicates that the system wants to phase separate. On the other hand, the non-symmetric MPS consistently give a positive $\Delta_{qp}$ for all non-zero $t_1$ values. In addition, we find that for $t_1 \gtrsim 0.02$, most non-symmetric MPS produce a lower ground state energy already at significantly smaller bond dimensions. We show some values in Table~\ref{tab:ecom} for comparison.  

\begin{table}
    \centering
    \begin{tabular}{c||c|c|c|c|c}
    \hline
        $t_1$ & 0.01 & 0.02 & 0.04 & 0.012 & 0.2 \\ \hline
        $D = 4800$ w/ SU(2) & -3.13045 & -3.14022 & -3.16512 & -3.27058 & -3.38192\\
        $D = 400$ w/o SU(2) & -3.13025 & -3.14058 & -3.16606 & -3.27410 & -3.38839\\
    \hline
    \end{tabular}
    \caption{Energy comparison between simulations with and without SU(2) symmetry}
    \label{tab:ecom}
\end{table}

\begin{figure}[h]
    \centering
    \includegraphics[width=\linewidth]{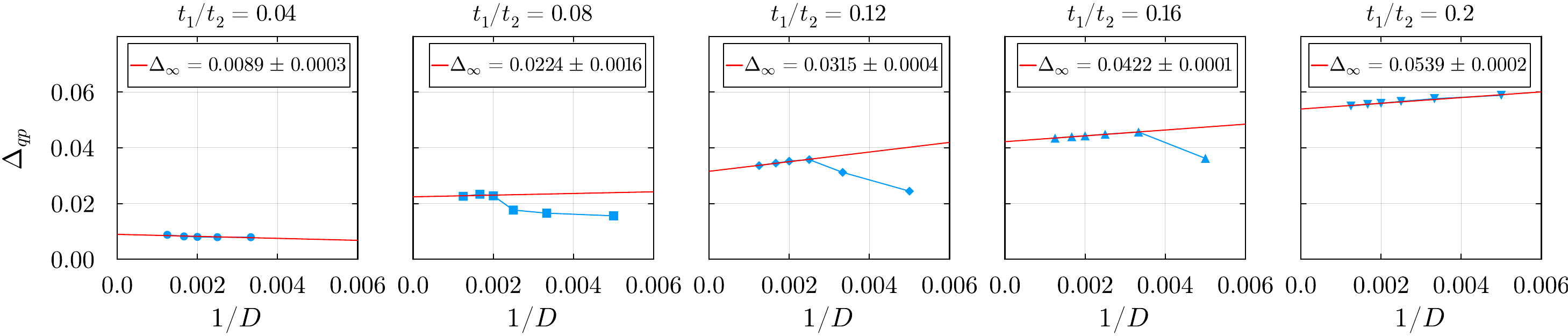}
    \caption{Extrapolations of the charge gap obtained from quasi-particle dispersion relations calculated without SU(2) symmetry being enforced in the MPS.}
    \label{fig:qpgap}
\end{figure}

To make sure that the positive $\Delta_{qp}$ in the non-SU(2)-symmetric case is not a finite-$D$ effect, we extrapolate $\Delta_{qp}$ to infinite bond dimension, as shown in Fig.~\ref{fig:qpgap}. For smaller $t_1$ we find stronger SU(2) breaking at smaller bond dimension, hence $\Delta_{qp}$ is not monotonic as $D$ increases. We only use the linear part at higher $D$ for the extrapolation. With this procedure we find $\Delta_{\infty}$ to be positive for all non-zero $t_1$ values.

\subsection{Translation symmetry breaking}
\begin{figure}[h]
    \centering
    \includegraphics[width=0.5\linewidth]{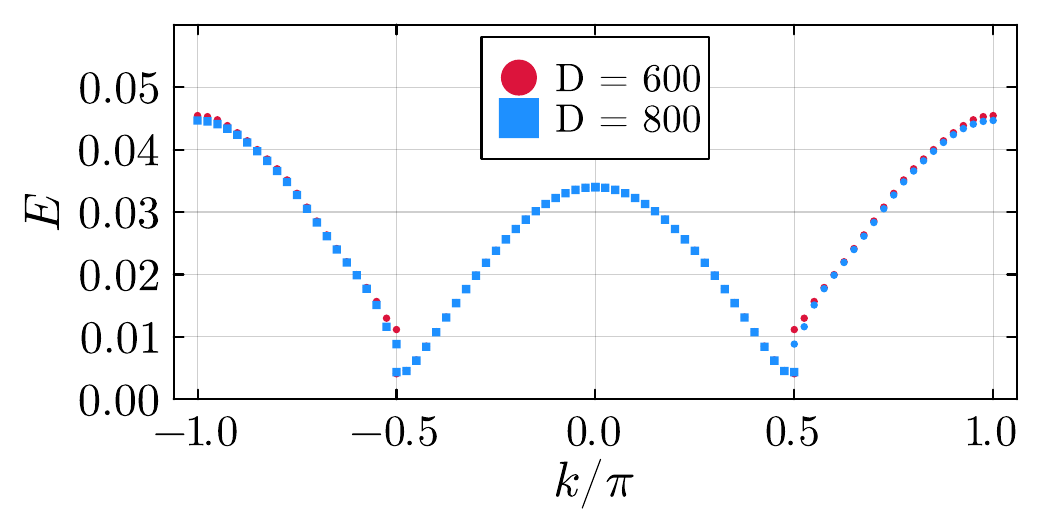}
    \caption{Weak translation symmetry breaking in the electron quasiparticle dispersion at $t_1 = 0.04$ with $D = 600/800$.}
    \label{fig:E6v8}
\end{figure}

Due to the doubling of the unit cell in our MPS simulations, we need to unfold the quasi-particle dispersion relations to represent the original Brillouin zone physics. Reflection symmetry implies that $E_n(k) = E_n(-k)$, and single unit-cell translation symmetry implies $E_1(k) = E_2(k+\pi)$, such that $E_1(\pm\pi/2) = E_2(\pm\pi/2)$ should hold for the numerically obtained dispersion relations in the folded zone (note that we work in units of the original unit cell spacing). However, we find that there is weak translation symmetry breaking, as can be seen in Fig.~\ref{fig:E6v8} from the small gap between $E_1(\pm\pi/2)$ and $ E_2(\pm\pi/2)$. This gap is only noticeable at small $t_1$, where the quasi-particle bands become very flat. We expect the gap at the folded zone boundary to be a finite-$D$ effect, as it decreases for larger $D$ as shown in Fig.~\ref{fig:E6v8}. It is numerically challenging to reach sufficiently high $D$ to see this gap actually close, as calculating the electron dispersion relation at $D=800$ without SU(2) symmetry already took 20 days with our computational resources.

\section{Critical properties of the spiral spin liquid}
\begin{figure}
    \centering
    \includegraphics[width=0.5\linewidth]{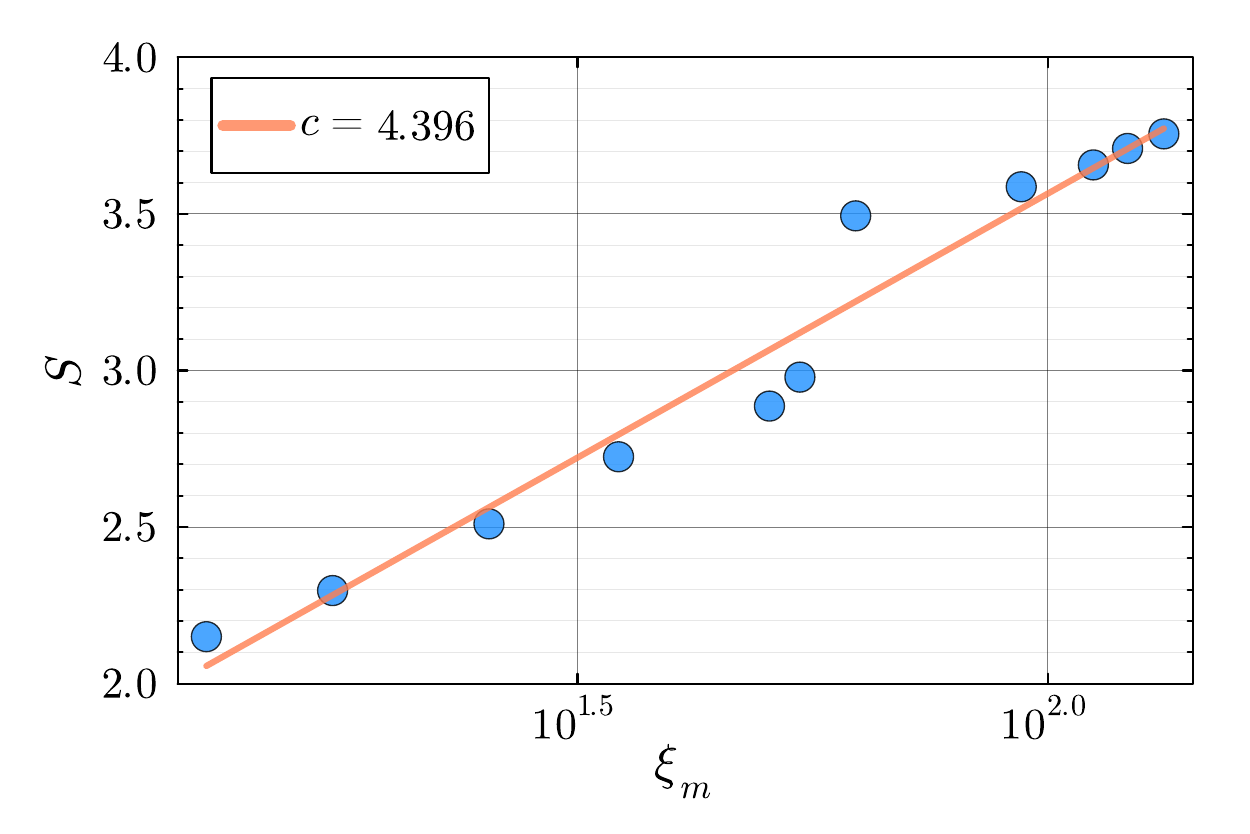}
    \caption{Entanglement entropy $S$ scaling with correlation length $\xi_m$ for the sawtooth spin model at $J_H/J_K = 2$. Data points are obtained with $D = [100, 200, 400, 800, 1600, 2000, 2400, 3200, 4000, 4800, 5600]$.}
    \label{fig:Svxi}
\end{figure}

To characterize the spiral spin liquid, our first attempt is again to use the CFT relation $S = \frac{c}{6}\log\xi_m + \text{const}$. As shown in Fig.~\ref{fig:Svxi}, the fitting is significantly worse than in the $SU(2)_1$ spin liquid. This is partly because at high values of $J_H/J_K$, the apex spins are only screened by the Heisenberg spin liquid at energy scales of order $\mathcal{O}(J_K^2/4J_H)$, and this happens entirely via quantum fluctuations (as argued above, at the classical level the apex spins are completely free). As a result, the spiral spin liquid has large entanglement, and at the highest bond dimension accessible to us ($D = 5600$) the correlation length $\xi_m$ is only $132$.

\begin{figure}
    \centering
    \includegraphics[width=\linewidth]{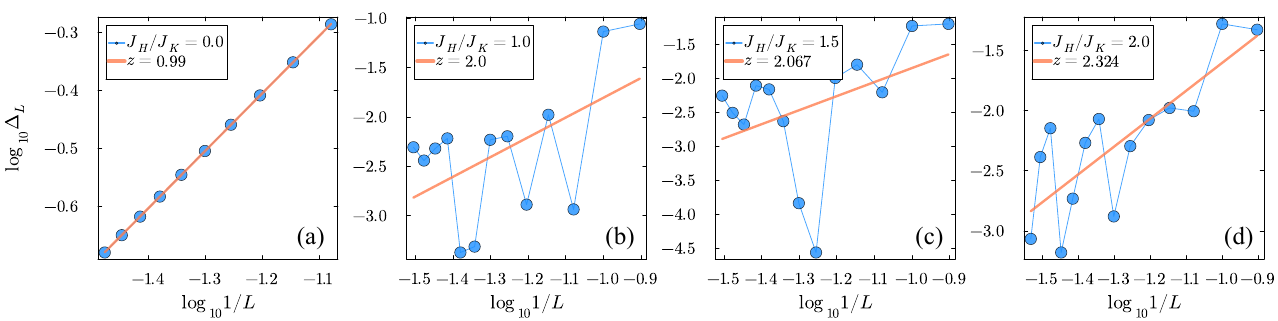}
    \caption{Log-log plot of the finite-sized gap $\Delta_L$ and system size $L$ in units of unit cells, obtained with ED on periodic boundary conditions. The dynamical exponent $z$ is fitted according to $\Delta_L \propto L^{-z}$.}
    \label{fig:finite-sizeGap}
\end{figure}

Due to the poor quality of the central charge fit and the unreasonably high value for $c$ obtained from it, we suspect a non-CFT critical point. We perform ED calculations with HPhi \cite{KAWAMURA2017180, IDO2024109093} for the finite-size gap $\Delta_L$ up to 34 spins with periodic boundary conditions. In Fig.~\ref{fig:finite-sizeGap}, we show fits to the relation $\Delta_L \propto L^{-z}$ in order to obtain the dynamical exponent $z$. At the Heisenberg point, panel (a) shows the standard $z \approx 1$. In contrast, panels (b-d) obtained in the spiral spin liquid show strong deviations from $z = 1$. When $J_H/J_K = 2$, where we know the ($c$-electron) spiral is commensurate with $k_0 = \pi/2$, there is some structure with period 4. However, from each set of points in the period of 4 we obtain a different value of $z$. In passing, we also note that for the spiral spin liquid the finite-size gaps are an order of magnitude smaller than for the Heisenberg chain at the same $L$. 

\section{Anderson model}

In this final appendix we sketch how to obtain the Kondo lattice model studied in the main text starting from an Anderson model. The Hamiltonian we consider consists of three parts
\begin{eqnarray}
H & = & H_c + H_p+ H_{c-p}
\end{eqnarray}
The first part is simply the kinetic term for the $c$-electrons, identical to the one in the Kondo lattice model in the main text:
\begin{equation}
H_c =  -t_1 \sum_{j,\sigma} c^\dagger_{j,A,\sigma}c_{j,B,\sigma} - t_2 \sum_{j,\sigma} c^\dagger_{j,B,\sigma}c_{j+1,A,\sigma} + h.c.
\end{equation}
The second term describes two electrons in a $p_x$ and $p_z$ orbital in every unit cell:
\begin{equation}
H_p = U \sum_j \sum_{\alpha=p_x,p_z} n_{j,\alpha,\uparrow} n_{j,\alpha,\downarrow} + V\sum_j n_{j,p_x}n_{j,p_z} -J \sum_j \mathbf{S}_{j,p_x}\cdot \mathbf{S}_{j,p_z} - \mu_p \sum_j (n_{j,p_x} + n_{j,p_z})
\end{equation}
Here we have defined
\begin{eqnarray}
n_{j,\alpha,\sigma} & = & c^\dagger_{j,\alpha,\sigma}c_{j,\alpha,\sigma} \\
n_{j,\alpha} & = & \sum_\sigma n_{j,\alpha,\sigma} \\
\mathbf{S}_{j,\alpha} & = & \frac{1}{2} c^\dagger_{j,\alpha}\boldsymbol{\sigma}c_{j,\alpha}
\end{eqnarray}
We take the parameters $U,V$ and $\mu_p$ such that there is one electron in the $p_x$ orbital, and one in the $p_z$ orbital, with a big gap to empty or doubly-occupied states. $J$ is the conventional Hund's coupling favoring a spin-$1$ state of the two electrons, which we take to be large.

Finally we consider $H_{c-p}$, which couples the electrons in the $p$-orbitals to the itinerant electrons. If we take the chain of $c$-electrons to be oriented along the $x$-direction, then reflection symmetry dictates that the coupling should take the form
\begin{equation}
H_{c-p} = -t_{p_x} \sum_j c^\dagger_{j,p_x}\left( c_{j,A} - c_{j,B} \right) -t_{p_z} \sum_j c^\dagger_{j,p_z}\left( c_{j,A} + c_{j,B} \right)  + h.c.
\end{equation}
If we take $t_{p_x} = t_{p_z}$, then the usual Schrieffer-Wolff transformation to go from the Anderson model to the Kondo lattice model produces following Kondo couplings:
\begin{align}
& \frac{J_K}{2} \sum_j \left[\mathbf{S}_{j,p_x}\cdot\left(c^\dagger_{j,A} - c^\dagger_{j,B}\right)\boldsymbol{\sigma} \left(c_{j,A} - c_{j,B}\right) + \mathbf{S}_{j,p_z}\cdot\left(c^\dagger_{j,A} + c^\dagger_{j,B}\right)\boldsymbol{\sigma} \left(c_{j,A} + c_{j,B}\right) \right] \\
& = \frac{J_K}{2} \sum_j \left[ \left(\mathbf{S}_{j,p_z} + \mathbf{S}_{j,p_x} \right)\cdot \left(c^\dagger_{j,A}\boldsymbol{\sigma}c_{j,A} + c^\dagger_{j,B}\boldsymbol{\sigma}c_{j,B} \right)  + \left(\mathbf{S}_{j,p_z} - \mathbf{S}_{j,p_x} \right)\cdot \left(c^\dagger_{j,A}\boldsymbol{\sigma}c_{j,B} + c^\dagger_{j,B}\boldsymbol{\sigma}c_{j,A} \right) \right]
\end{align}
When the Hund's coupling is very large we can project this coupling term into the spin-$1$ subspace for the two electrons in the $p$-orbitals. Since this subspace corresponds to the symmetric subspace, the projection annihilates the term containing $\mathbf{S}_{j,p_z} - \mathbf{S}_{j,p_x}$. We can also replace $\mathbf{S}_{j,p_z} + \mathbf{S}_{j,p_x}$ with the spin-$1$ operator $\mathbf{S}_{j}$. The effective Kondo Hamiltonian thus becomes 
\begin{equation}
H = -t_1 \sum_{j,\sigma} c^\dagger_{j,A,\sigma}c_{j,B,\sigma} - t_2 \sum_{j,\sigma} c^\dagger_{j,B,\sigma}c_{j+1,A,\sigma} + h.c. + \frac{J_K}{2}\sum_j \mathbf{S}_j \cdot \left(c^\dagger_{j,A}\boldsymbol{\sigma}c_{j,A} + c^\dagger_{j,B}\boldsymbol{\sigma}c_{j,B} \right)
\end{equation}
This is almost the Kondo lattice model studied in the main text, except that the local moments have spin $1$ instead of spin $1/2$. To remedy this, we add an additional spin $1/2$ in every unit cell, which we strongly couple via an anti-ferromagnetic exchange to the spin $1$. The model in the main text also contains an exchange coupling between the local moments. This can either come from the chain of spin-$1/2$'s that we couple to the above spin-$1$ Kondo lattice model, or it can also naturally be induced for the electrons in the $p_x$ orbitals, as these are oriented parallel to the chain and can therefore generate a significant super-exchange between neighboring unit cells.

\end{document}